\documentclass[10pt]{iopart}
\usepackage{graphicx} 
\usepackage{multirow}       
\usepackage{booktabs}
\usepackage{cite}
\usepackage{textcomp}
\usepackage{gensymb}
\graphicspath{{Figures/}}
\begin{document}

\title{Hermetic chip-scale packaging using Au:Sn eutectic bonding for implantable devices}
\author{Katarzyna M. Szostak$^1$, Meysam Keshavarz$^2$, and Timothy G. Constandinou$^{1,3}$}
\address{$^1$ Next Generation Neural Interfaces (NGNI) Lab, Imperial College London, Bessemer Building, London SW7~2AZ, UK}
\address{$^2$ The Hamlyn Centre, Imperial College,  Bessemer Building, London SW7~2AZ, UK}
\address{$^3$ UK Dementia Research Institute, Imperial College,  South Kensington Campus, London SW7~2AZ, UK}
\ead{k.szostak@imperial.ac.uk}
\vspace{12pt}

\begin{abstract}
Advancements in miniaturisation and new capabilities of implantable devices impose a need for the development of compact, hermetic, and CMOS-compatible micro packaging methods. Gold-tin-based eutectic bonding presents the potential for achieving low-footprint seals with low permeability to moisture at process temperatures below 350\degree C.  In this paper, a method for the deposition of gold-tin eutectic alloy frames via sequential electroplating from commercially available solutions, with no special fabrication process, is described in detail. Bond quality was characterised through shear force measurements, scanning electron microscopy, visual inspection, and immersion tests. Characterisation of seals geometry, solder thickness, and bonding process parameters were evaluated, along with toxicity assessment of bonding layers to the human fibroblast cells. With a successful bond yield of over 70\% and no cytotoxic effect, AuSn eutectic bonding appears as a suitable method for the protection of integrated circuitry in implantable applications.
\end{abstract}
\vspace{2pc}
\noindent{\it Keywords}: AuSn, Eutectic, Bonding, Hermeticity, Micropackaging, Encapsulation, Reliability, Implant

\section{Introduction}

One of the most critical design challenges in the growing sector of miniaturised, smart implantable devices is providing reliable packaging of a small footprint and good hermeticity.  \cite{chong2020non, ahn2019emerging}. Novel implantable devices ever-expanding capabilities more often include fully wireless communication and on-node data processing, thus imposing more rigorous requirements for device packaging~\cite{musk2019integrated}. Nowadays, implant packages, aside from ensuring small size, reliability, and biocompatibility, may additionally require CMOS compatibility and transparency to the communication signals~\cite{das2020biointegrated, seymour2017state, yang2020review}. Until now, the majority of implants were packaged by encasing entire systems in metal or glass cases or by using thick outer coatings, mainly made of medical-grade silicones~\cite{ahn2019emerging, szostak2017neural, yang2020review}. Typically, these solutions either do not provide an appropriate level of hermeticity or are too large and not fit for wireless systems with integrated electronics. At the same time, it is not certain whether standard IC insulation layers alone are sufficient in providing enough protection against the inherently wet and often corrosive character of the tissue environment over a long time~\cite{maloney2005vivo, vanhoestenberghe2013corrosion, lamont2021silicone}. Recently, approaches of using thin, ALD-deposited high-K ceramic materials such as Hafnium Dioxide and multilayers of Silicon Dioxide with Hafnium Dioxide as a protective coating on neural implants have emerged \cite{jeong2019conformal, shen2020ceramic}. They hold the promise of thin, stable coatings; however, they would not solve the issue of mechanical stability of chips stressed by the tissue nor integration difficulty of more advanced microelectrode designs.

For that reason, the prospect of hermetically enclosing implant electronics at a wafer- or chip-scale by using narrow seal rings with the addition of in-silicon feedthroughs is highly desirable~\cite{liu2020bidirectional}. 

From a myriad of bonding techniques available in electronics and MEMS packaging, only a few do not require high-temperature processing nor contain toxic materials while providing good seal hermeticity, thus making them applicable for implantable devices containing embedded ICs~\cite{pelzer2005wafer, esashi2008wafer}. To list a few, anodic bonding methods require non-CMOS-compatible high electrostatic fields, thermocompression, and glass frit bonding use very high temperatures, while the majority of solders contain copper and silver proven to be cytotoxic elements.  Amongst all packaging methods, eutectic bonding is well-established and can be performed using relatively relaxed environmental parameters, while the availability of several different eutectic compositions enables the choice of the one that would suit the application needs the best. A clear advantage of metal-based sealing is the tolerance to substrate unevenness and very low permeability to moisture, superior to even those of silicon or glass~\cite{morales2015evaluating}. \newline

\begin{figure} [h]
    \centering
    \includegraphics[width=1\linewidth]{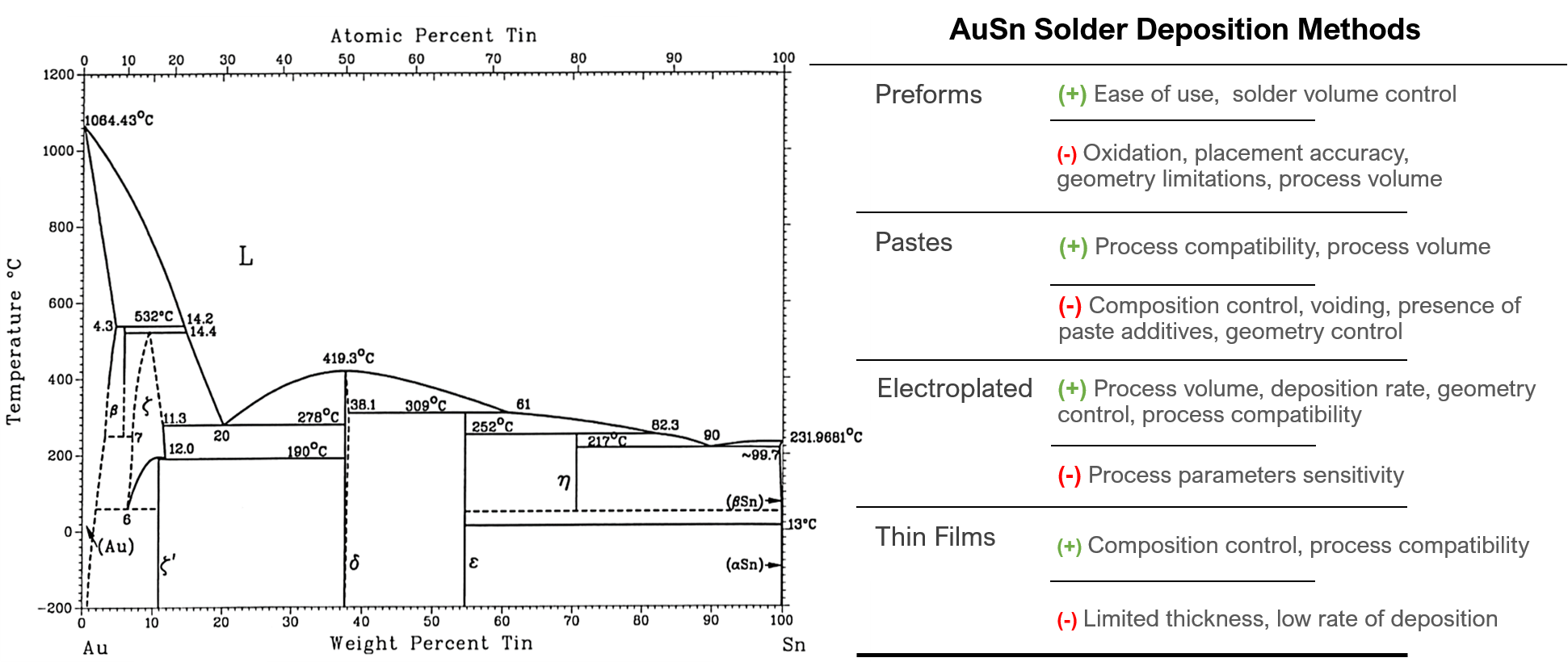}
    \caption{Left: Au-Sn binary phase diagram \cite{okamoto1984sn}. Right: most common AuSn solder deposition methods and their characteristics.}
    \label{phasediag}
\end{figure}
While primarily used for die attach and flip-chip bonding in semiconductor and optoelectronics packaging, eutectic bonding can be also used for hermetic sealing~\cite{saeidi2013technology, Overmeyer2014}. The concept relies on using a single solder frame at the interface of two substrates protecting circuitry and devices enclosed within. Since the joint’s thickness and width are in the range of tens of microns or less, considerable design freedom and reduction in the package size can be achieved. Out of a number of eutectic compositions existing, the one particularly interesting for the application in implantable technologies is gold-tin metallurgy~\cite{zhou1999sn, lee1991new}. This is thanks to Au-Sn favourable characteristics of high joint strength, fluxless processing, reliability, and moderate liquidus temperature of 278\degree C.  

Although Au-Sn bonding has been already demonstrated for the packaging of MEMS devices, particularly for RF and inertial applications, in order to utilise this method more widely and effectively, more studies on bond characteristics and reliability are needed (Table.~\ref{table_applications})~\cite{zoschke2013hermetic, durante2011reliable, giudice2013sn}.  To deliver a good quality seal, it is of much importance to control the deposition of eutectic composition correctly, as at precisely 80:20 AuSn ratio by weight (Fig.~\ref{phasediag}), the low eutectic temperature of the alloy is paired with desirable properties of Au$_5$Sn intermetallic phase~\cite{tsai2005controlling, szostak2018hermetic}. There are several possible methods for depositing gold-tin alloy (Fig.~\ref{phasediag}), whereas the most popular and cost-effective approach is electrochemical deposition.  This is typically achieved through alloy plating, which suffers problems of short service life and solution stability~\cite{mcnulty2008processing}. To overcome that issue, composition plating can be executed by sequential electroplating, which involves using different chemical solutions to deposit each component of the eutectic alloy~\cite{szostak2018hermetic}.    

In this study, the effectiveness of tin-rich sequentially deposited Au-Sn alloy eutectic bonding is evaluated to determine critical parameters for achieving high-quality, hermetic seals. The influence of bond geometry, solder thickness and reflow parameters on bond mechanical properties and hermeticity are assessed. To establish the preliminary validation of biocompatibility of the method in implantable applications, the cytotoxicity of plating chemistry and the alloy itself were assessed by culturing human fibroblast cells on reflown samples.

\begin{table} [h]
\label{table_applications}
\caption{Selected applications of AuSn solder rings for hermetic sealing.}
\footnotesize
\begin{tabular}{@{}llc@{}}
\toprule
\multicolumn{1}{c}{\textbf{Application:}}                                          & \multicolumn{1}{c}{\textbf{Details:}}                                                                                                                        & \textbf{Year}                                       \\ \midrule
\begin{tabular}[c]{@{}l@{}}Smart inertial MEMS\\ sensor systems\end{tabular}       & \begin{tabular}[c]{@{}l@{}}- Chip-to-wafer integration\\ - AuSn frame: 18\,\textmu m height, electroplated              \\ - Bonding: 300\degree C, 6\,kN force\end{tabular} & \begin{tabular}[c]{@{}c@{}}2008\\ \cite{marenco2008vacuum}\end{tabular} \\ \midrule
RF MEMS switch                                                                     & \begin{tabular}[c]{@{}l@{}}- Wafer-level integration\\ - AuSn frame: 5\,\textmu m height, electroplated\\ - Bonding: 310\degree C\end{tabular}                              & \begin{tabular}[c]{@{}c@{}}2010\\ \cite{ferrandon2010hermetic}\end{tabular} \\ \midrule
\begin{tabular}[c]{@{}l@{}}Piezoelectric silicon\\ resonators\end{tabular}         & \begin{tabular}[c]{@{}l@{}}- Wafer- and array-level integration\\ - AuSn frame: 4\,\textmu m height, electroplated\end{tabular}                                      & \begin{tabular}[c]{@{}c@{}}2011\\ \cite{durante2011reliable}\end{tabular} \\ \midrule
\begin{tabular}[c]{@{}l@{}}MEMS microresonators\\ and microbolometers\end{tabular} & \begin{tabular}[c]{@{}l@{}}- Chip-level integration\\ - AuSn frame: 20\,\textmu m, electroplated\\ - Bonding: 340\degree C\end{tabular}                                     & \begin{tabular}[c]{@{}c@{}}2013\\ \cite{giudice2013sn}\end{tabular} \\ \midrule
MEMS timing devices                                                                & \begin{tabular}[c]{@{}l@{}}- Wafer-level integration\\ - AuSn frame: 19\,\textmu m, electroplated\\ - Bonding: 300\degree C\end{tabular}                                    & \begin{tabular}[c]{@{}c@{}}2013\\ \cite{zoschke2013hermetic}\end{tabular}
\end{tabular}
\end{table}

\section{Materials and Methods}
\begin{figure} 
    \centering
    \includegraphics[width=1\linewidth]{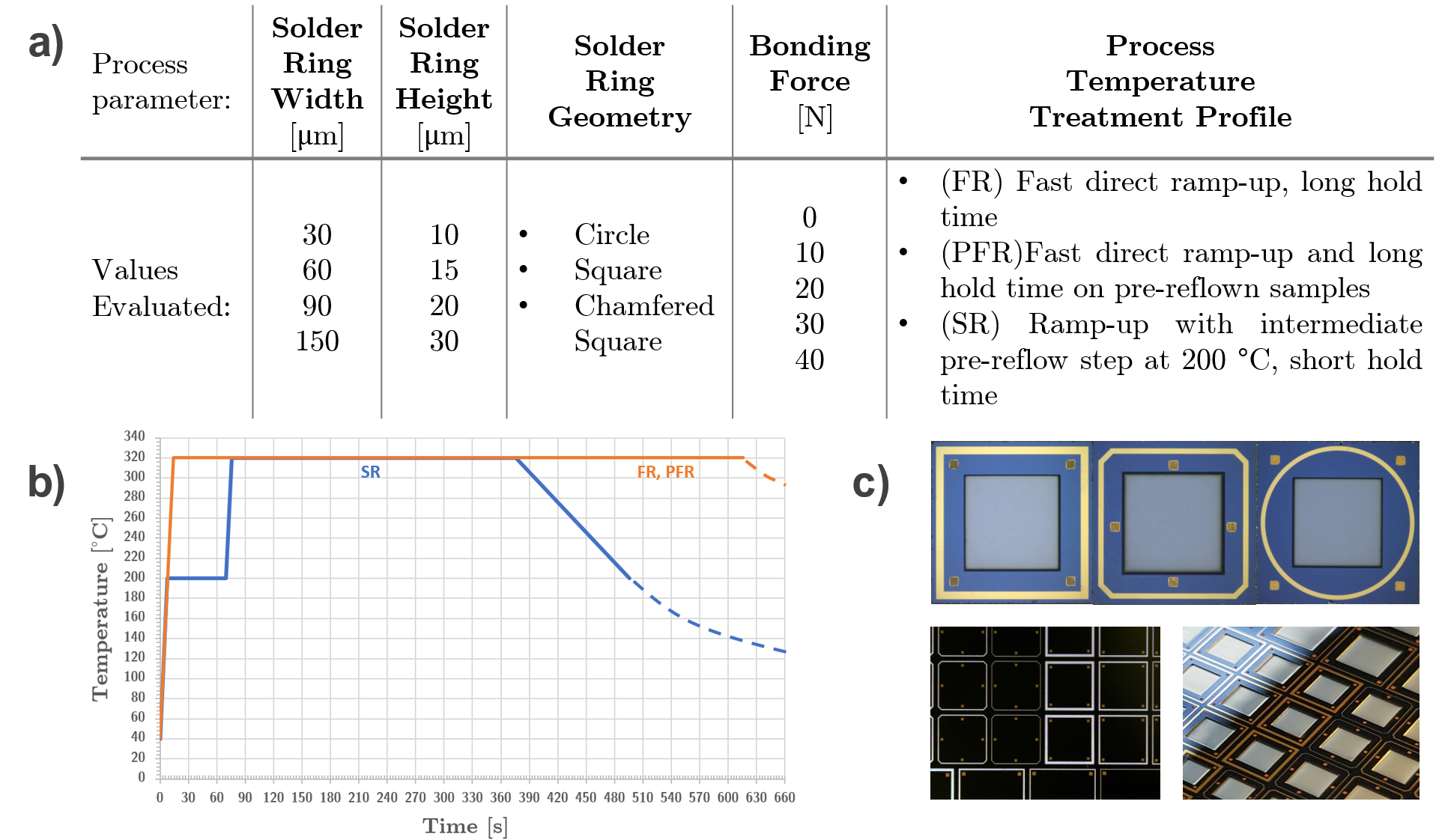}
    \caption{A) An overview of process parameters tested in the study B) Reflow temperature profiles evaluated C) Evaluated seal geometries of the square, chamfered square, and round frames.}
    \label{parameters}
\end{figure}

All experiments were carried out in an academic cleanroom laboratory setting (Centre for Bio-Inspired Technology Cleanroom, Imperial College London and London Centre for Nanotechnology Cleanroom, University College London) using 4-inch, double side polished, 525\,\textmu m thick (100) silicon wafers of total thickness variation lesser than 2.3\,\textmu m. Substrates were processed in two batches, separately for lid and base die samples. All mask layers were custom designed, so that cap wafers sealing rings match with those on the base wafers. Overall, two sample sizes (4\,mm$^2$, 9\,mm$^2$), with four widths (30\,\textmu m, 60\,\textmu m, 90\,\textmu m, 150\,\textmu m) and three different shapes (circular, rectangular, and chamfered) of seal rings have been tested. Eutectic stacks heights were electroplated to the heights of either 10\,\textmu m or 15\,\textmu m, which corresponds to Au(6\,\textmu m)- Sn(4\,\textmu m) and Au(9\,\textmu m)-Sn(6\,\textmu m) tin-rich compositions, respectively. The range of tested parameters was based on the previous literature (Tab.~\ref{table_applications}) of the subject and project's constraints regarding the package size, which was to be kept smaller than 4$\times$4\,mm$^2$. Additionally, some of the geometries contained gold-only rectangular standoffs, placed either inside or outside of the seal rings (Fig.~\ref{parameters}). 

\subsection{Fabrication of base and lid wafers}
\begin{figure} 
    \centering
    \includegraphics[width=1\linewidth]{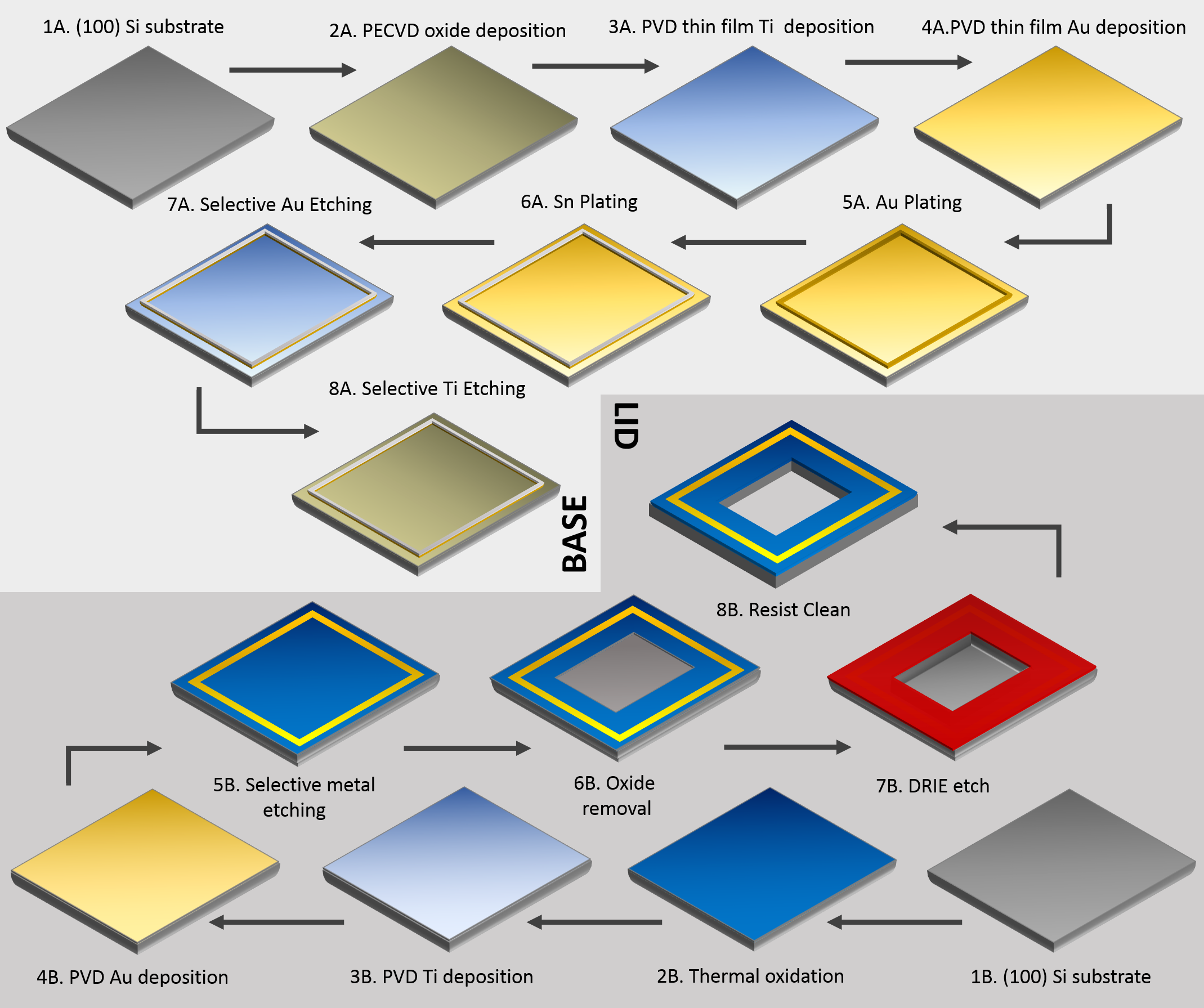}
    \caption{An overview of base (above) and lid (below) wafers fabrication process flow}
    \label{pflow}
\end{figure}

A simplified fabrication process flow is depicted in Fig.~\ref{pflow}. Processing commenced with complete RCA clean, after which lid wafers were oxidised with approximately 320\,nm dry thermal silicon oxide and base wafers were coated with approximately 960\,nm PECVD-deposited oxide (Mesc Multiplex 13241 system, STS, UK). The difference in oxide thickness was chosen, so that base wafers' oxide thickness corresponds better to those on the BEOL layers used in CMOS chips. Next, adhesion layers of either Cr or Ti were sputtered (PVD75, K.Lesker, USA) to the thickness of 20\,nm and directly followed by sputter deposition of 100\,nm layer of Au, which would work as a seed layer for later electrodeposition steps.  Subsequently, selected substrates were photolithographically patterned with 5\,\textmu m AZ 15nXT (115\,CPS) photoresist to form an array of square-shaped openings, which were then electroplated with gold to the height of 5\,\textmu m using parameters described in the later steps. This is to create gold standoffs near the solder lines, which would work as possible spillage stoppers preventing total solder collapse during bonding. Then, the photoresist was stripped in heated MICROPOSIT Remover 1165, and substrates were cleaned in Acetone and Isopropanol baths, followed by DI water rinse and nitrogen dry.

Until this point, substrates intended for both base and lid of the package were processed the same. In the next step, lid wafers were photolithographically patterned with the shapes of seal rings in AZ1518 HS photoresist followed by differential etching away of previously sputtered Ti/Au layers and photoresist removal in acetone. In the next step, lid wafers were again patterned using SPR-220 photoresist to form central square openings used for the formation of central cavities. These cavities provide space for the placement of cut copper film pieces (in the future to be replaced by humidity sensors) used to evaluate bond hermeticity.  After removing silicon dioxide in buffered hydrofluoric acid, cavities were etched down in the Bosch process to the depth of 200 (Multiplex Pro DRIE, SPTS, UK). Then, lid wafers were stripped of photoresist, solvent- cleaned, and diced into 4\,mm$^2$ and 9\,mm$^2$ dies (DAD3230 dicing saw, DISCO Corporation, Japan). 

After electroplating spillage-stoppers, base wafers were sectioned into four pieces to separate wafer areas containing designs of different sizes. This was to avoid electroplating with starkly different local current densities across the wafer and thus allowing for better control of the deposition process.  Then, base wafer sections underwent another photolithography to pattern outlines of seal rings. For that, a thicker version of the previously used photoresist AZ 15nXT (450\,CPS) of the thickness of 120\% of desired plated deposit thickness was used. After that, substrates were sequentially electroplated with gold and tin layers. Once that was achieved, the photoresist was stripped, and wafers solvent cleaned. Next, samples underwent selective etching of thin films of adhesion and seed layers in gold, chromium, and titanium etchants (KI, HClO$_4$:(NH$_4$)$_2$[Ce(NO$_3$)$_6$], HF; respectively) thus leaving only layers of deposited solder on the silicon dioxide base. Finally, base wafer sections were diced to singulate dies of sizes corresponding to those diced from lid wafers, and exact solder heights were measured by contact profilometry. One set of samples was then annealed in a vacuum oven at 200\degree C for four hours. 

\subsection{AuSn Alloy Sequential Electrochemical Deposition}
\begin{figure} [h!]
    \centering
    \includegraphics[width=1\linewidth]{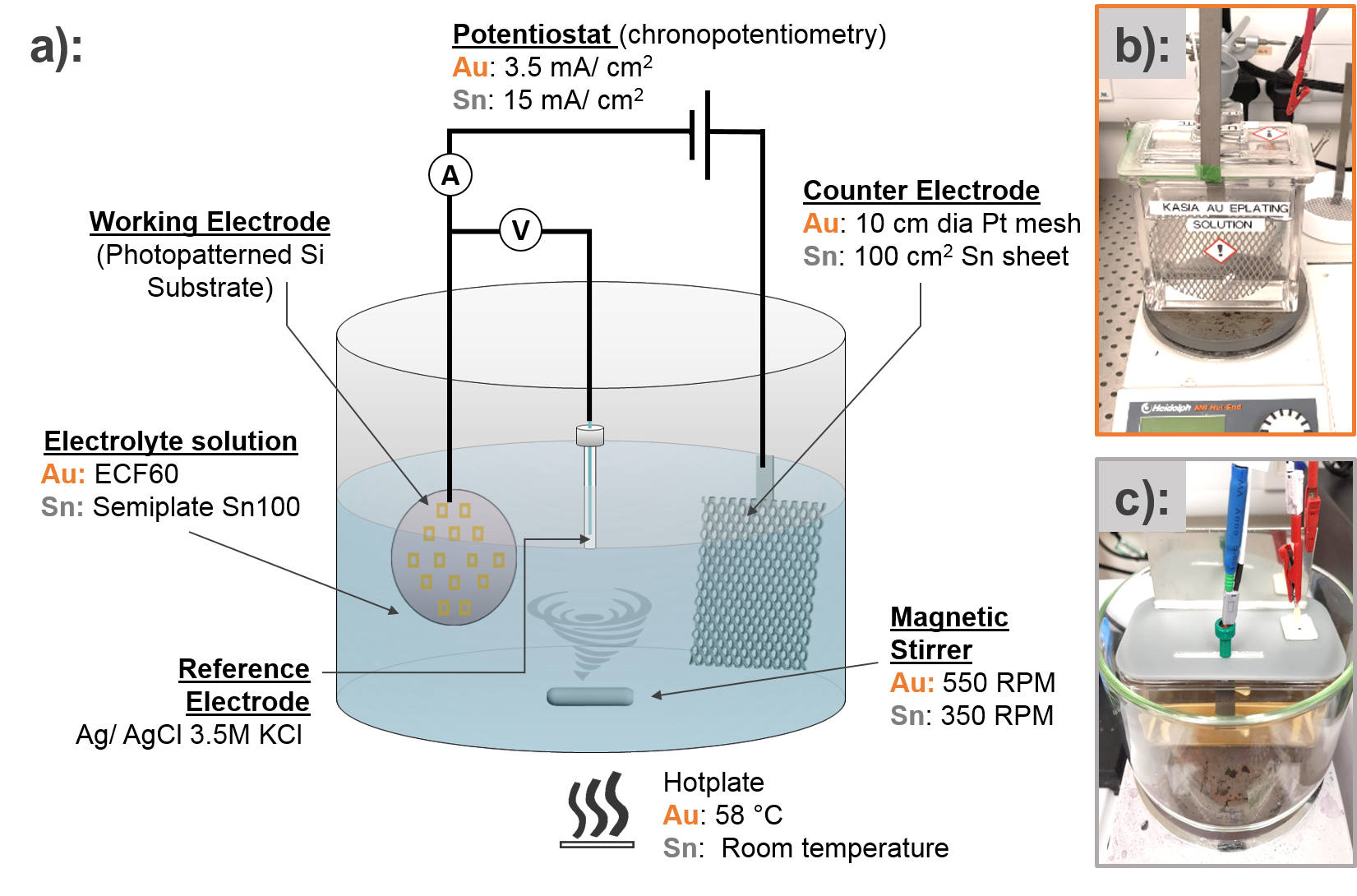}
    \caption{A) Sketch of setup used for electrochemical deposition of Au and Sn layers; B) Picture of Au-electroplating setup realisation; C) Picture of Sn-electroplating setup realisation.}
    \label{ecd_setup}
\end{figure}

For all described electroplating experiments, custom electrochemical deposition setup
was assembled, which is depicted in Fig.~\ref{ecd_setup}. An electrochemical workstation in a chronopotentiometry mode (Versastat-3-400, Princeton Applied Research, USA) was employed as a power source. To ensure better control over plating current, depositions were conducted in a three-electrode system, using double junction 3.5~KCl-filled silver/silver chloride (Ag/AgCl) electrode as a reference.  Electrical between working electrode terminal and samples was achieved via custom made spring-loaded contacts. The thickness of grown deposits was regularly monitored by measuring grown layer ex-situ using a stylus profilometer(Dektak~X, Veeco, USA).

For gold electrochemical deposition, a proprietary sulphite based ECF60 (Metalor, UK) gold bath, without additives or grain refiners, was used. This chemistry was chosen because sulphite-based Au-plating solutions are known to produce layers of ductile, soft pure gold with good throwing power and are safer to use than cyanide-based chemistry. Electroplating was carried out against a 10\,cm diameter platinised titanium mesh counter electrode at an operating temperature of 58\degree C. The exact plating current density value was chosen based on the manufacturer’s recommendations and previous literature on the subject at 3.5\,mA/cm$^2$ where the built-in stress of deposited gold layer is the lowest. To limit the formation of an ion-depleted solution layer in the cathode vicinity over time, the solution was continuously stirred by immersed magnetic pellet with a spin speed of 550\,rpm/s. Throughout the gold electroplating process, the counter electrode and samples were kept 5\,cm apart, the distance defined by the geometry of the plating vessel. 
Electroplating of tin was performed using proprietary methane sulphonic acid-based NB Semiplate Sn100 solution (Microchemicals GmbH, Germany), producing matte tin layers of high purity and low internal stresses. The deposition was performed using 100\,cm$^2$ 99.9\% tin sheet counter electrode at room temperature.  The plating solution was constantly stirred with 350\,rpm/min speed using a magnetic pellet to disperse hydrogen gas bubbles forming during deposition and aid the fresh solution's flow to the sample’s surface. Counter and working electrodes were kept at a constant distance of 3.5\,cm from each other, a gap defined by the geometry of the plating vessel. 

\subsection{Bonding process}
\begin{figure} 
    \centering
    \includegraphics[width=1\linewidth]{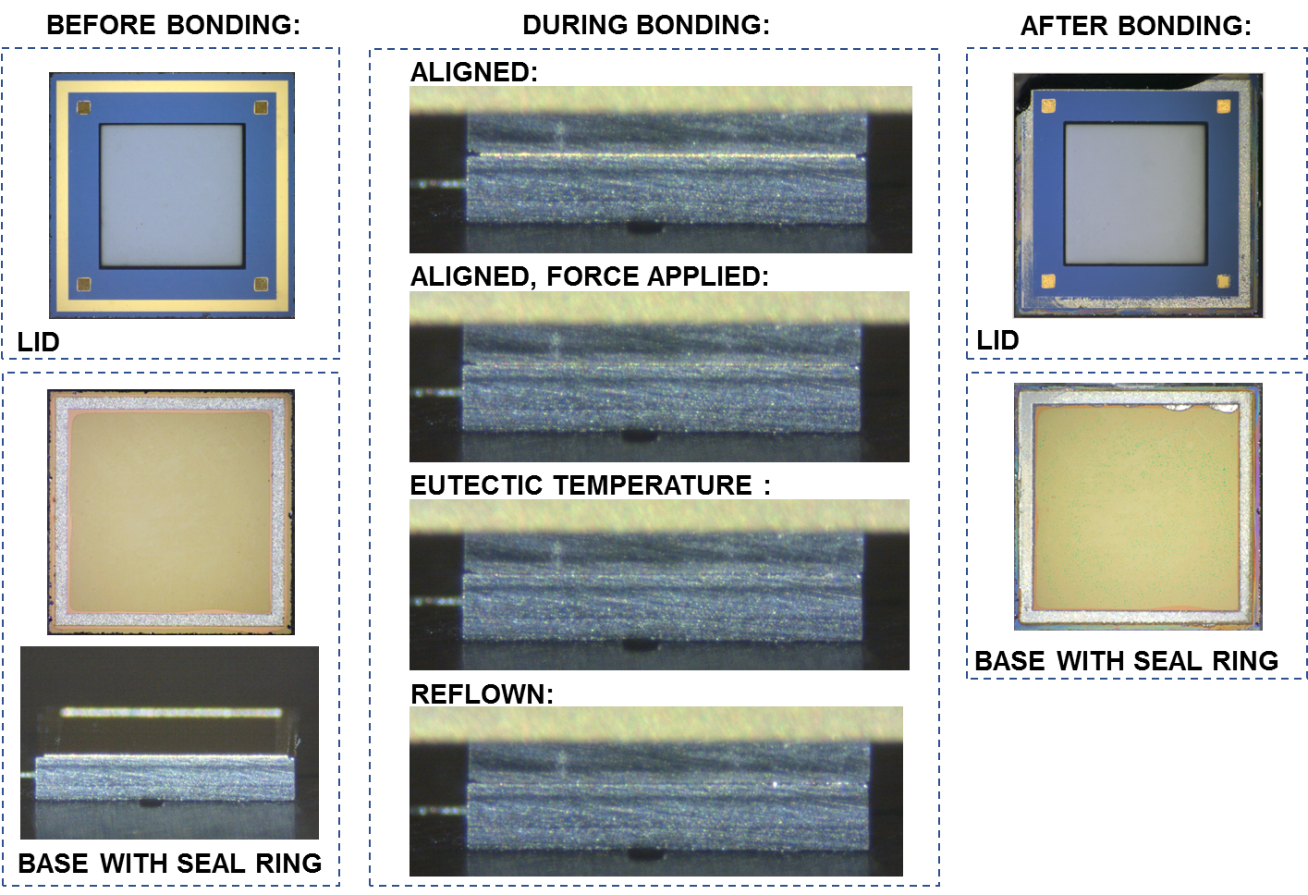}
    \caption{Microscope photographs of lid and base chips’ surface as seen before bonding (left) and after reflow and following shear tests (right). The side profile of the chip stack as seen during the bonding procedure (centre). Solder collapse and change in colour appearance of solder are observed.}
    \label{bonding}
\end{figure}

To test the effectiveness of sequential electrodeposition to form eutectic composition and evaluate bond characteristics, corresponding lid and base dies were bonded together using a die bonder (Lambda Fineplacer, Finetech, Germany). In total, 110 sample pairs were processed. All samples were bonded at least three weeks after the last electroplating step. Shortly before bonding, sample pairs were briefly cleaned in acetone and isopropanol and carefully dried with a nitrogen gun to remove any possible dicing-induced contamination. Several samples were additionally equipped with a small cut out of copper film placed inside the lid’s cavity to monitor the hermeticity of a formed seal. Lid and base dies were aligned at complete separation using bonders’ optical system, brought together to direct contact and subjected to the force of value ranging from 0\,N to 40\,N per solder ring. The bonding was done by applying two different temperature profiles, depending on the tested batch (orange and blue lines on temperature profile in Fig.~\ref{parameters}b). Batch PFR containing samples pre-reflown in the vacuum oven and batch FR were bonded by applying 320\degree C on the sides of both base and lid dies, with the ramp rate of 20\,K/s and held at the maximum temperature for 10\,min, after which heating was turned off, and the sample was allowed to return to room temperature without any external cooling. Samples from batch SR were reflown by applying 200\degree C to both base and lid dies, with the ramp rate of 20\,K/s for 60\,s followed by a further ramp-up to 320\degree C bonding temperature where they were held for 5\,min. After that, samples were first actively cooled to 200\degree C and then allowed to cool down without any external aids. Once the bonded pairs achieved room temperature, the force was removed, and samples were removed from the bonder for characterisation. Throughout the entire bonding process, samples were observed for the eutectic temperature, premature melting, spillages, and bond collapse via the side camera (Fig.~\ref{bonding}).

\subsection{Post-bonding characterisation}
The objectives for an implantable device package are the protection of the device against an external, humid environment of living and constantly mobile environment over a long time. Thus, the effectiveness of the bonding approach was characterised with a focus on shear strength, hermeticity, cell response to sealing material, and the overall yield of the bonding process. 
Hermeticity was qualitatively assessed by evaluating the change in surface oxidation of copper films as exposed to a wet environment in immersion tests. Of several leak detection methods available to use as a measure of hermeticity, the copper test was chosen, because gold-standard methods, such as the helium leak test, do not offer sufficient resolution for the device's sub-millimetre cavity size~\cite{vanhoestenberghe2011limits}. In the future, integration of an on-chip humidity sensor would provide a more quantitative measure of hermeticity~\cite{mazza2018integrated, akgun2020chip}. Directly after bonding, a sample set containing enclosed copper film pieces was placed in the water for 12 weeks along with the exposed copper control sample. The liquid was changed weekly. Then, samples were debonded and enclosed copper pieces evaluated for the change in appearance. The colour change of the copper film could indicate the progress of oxidation caused by moisture ingress inside a bonded sample. 

\begin{figure} 
    \centering
    \includegraphics[width=1\linewidth]{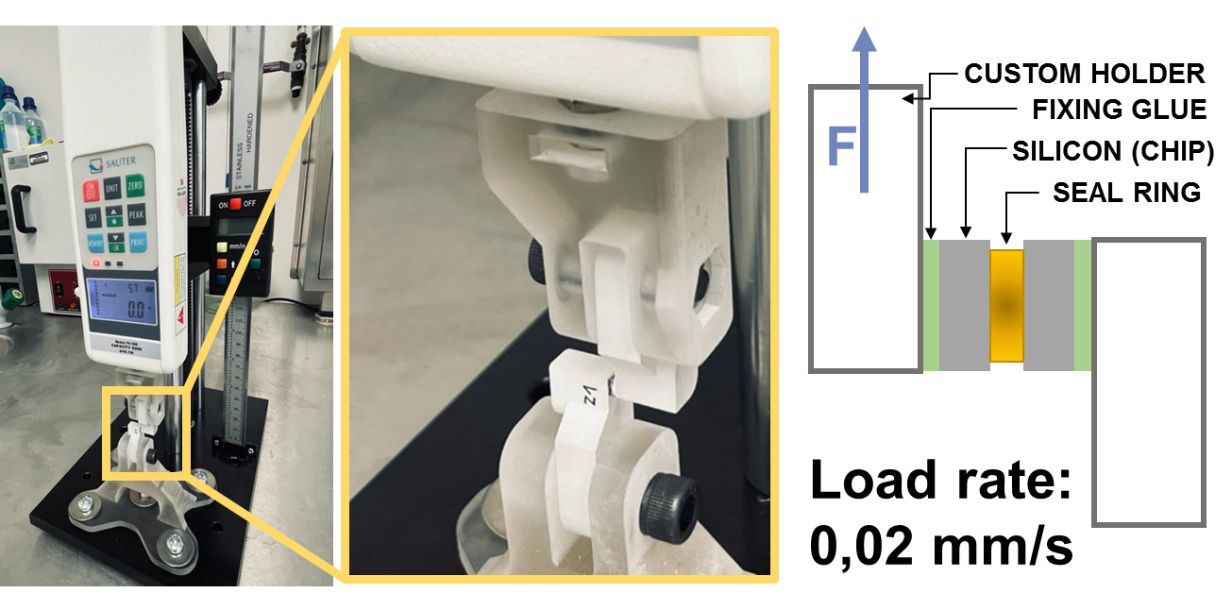}
    \caption{Shear test setup.}
    \label{shear}
\end{figure}

The strength of single seal frames was measured by mean of shear tests. The tests were carried out using force gauge (FH500, Sauter, Germany) fitted with custom 3D printed holders to which the backside surface of measured samples was carefully fixed using fast-setting adhesive (Fig.~\ref{shear}). The adhesive layer was thin enough to prevent it from bleeding into the interface between lid and base dies.  The load was applied with the approximate rate of 0.02\,mm/s until samples debonded at the interface and the maximum measured force was registered. The shear strength was then calculated by dividing the recorded force by the total bonded area. Samples were subsequently inspected with an optical microscope and scanning electron microscope to examine for the extent of possible spillages and evaluate the fracture mechanism of broken bonds. 

\subsection{Cell viability assessment}
Selected nine reflown and debonded samples were used as substrates for the human fibroblast cell viability and cytotoxic assessment. Samples used were sectioned off wafers before dicing; therefore, they were larger, each containing several seal rings. This was to evaluate the cytotoxic response of human fibroblast cell line to the AuSn solder in eutectic composition as deposited using sequential electroplating from sulphite and methane sulfonic acid-based gold and tin plating solution chemistries. Biocompatibility was assessed via analysis of cell morphology as inspected by scanning electron microscopy. Initially, samples were sterilised by exposing them to UV light for 20\,min followed by washing with PBS three times before adding to the cell culture medium. Next, cultured human fibroblast cells were employed in cell experiments to ascertain the cytotoxicity and cellular response to the fabricated layers. The cell line used in this experiment -fibroblast cell BJ (ATCC\textsuperscript{\textregistered} CRL-2522\textsuperscript{TM}), was purchased from ATCC and cultured in  Dulbecco’s modified Eagle’s medium (DMEM) containing 10\% heat-activated fetal bovine serum with 1\% penicillin-streptomycin antibiotics (Pen-strep). Subsequently, the cells were seeded and cultured on reflown AuSn substrates and placed in Petri dishes with a seeding density of 750,000~cells/~cm$^2$. The cells were incubated in the medium for 24, 48, and 72 hours at 37\degree C and 5\% CO$_2$ while they were examined every 12\,h.

\section{Results}
The majority of samples tested exhibited very good mechanical characteristics of produced joints. Across various geometries and parameters examined, the overall yield of the bonding process was 73\%.  Out of the total of 110 sample pairs processed, 24\% have produced seals with mechanical characteristics too weak to allow to conduct shear strength measurement and debonded upon normal handling.  Within the well-bonded group, 12\% of samples exhibited shear strength values below the minimum as described by MIL-STD-883G military standard for die-attach calculated per total bond area~\cite{inseto_uk_2020}, thus reducing the total yield of bonding process across the range of parameters tested to approximately 64\%.

\begin{figure} [h]
    \centering
    \includegraphics[width=1\linewidth]{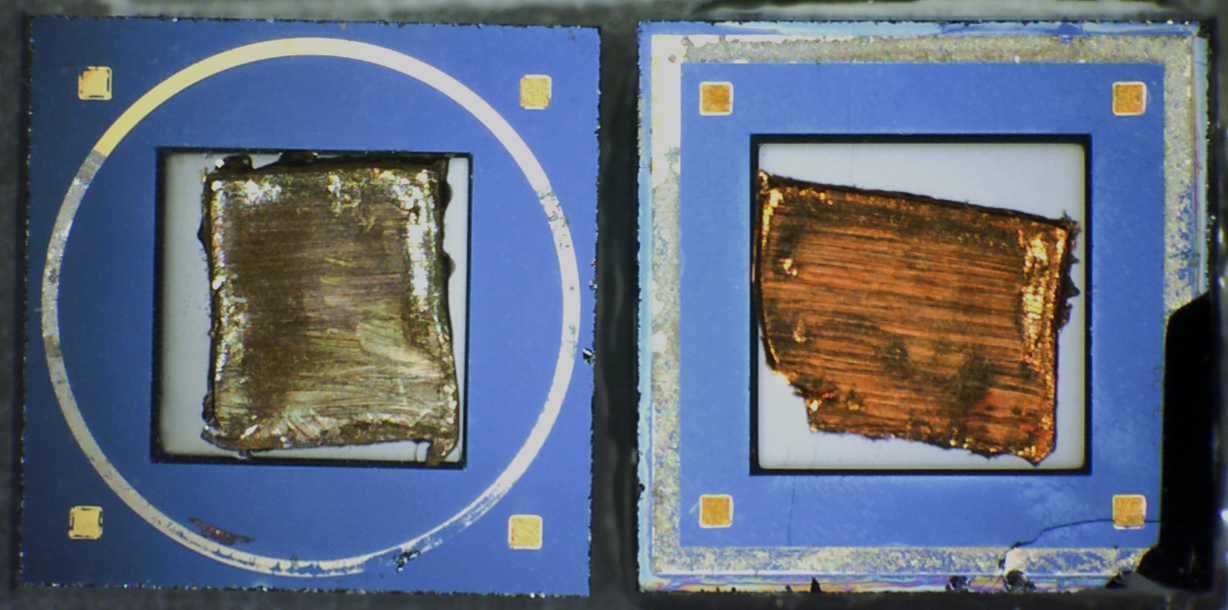}
    \caption{Comparison of an oxidation level of copper film immersed for 12 weeks in the water. Left: control sample (exposed to moisture) right: sample sealed by AuSn frame.}
    \label{copper}
\end{figure}

No correlation was observed between the shape of the seal lines and the resulting measured bond strength. For the frames of the same width, all shapes (circle, square, chamfered square) exhibited similar ranges of measured shear strength values.  Thus, when mechanical properties are of the foremost importance, the shape of solder rings are not a relevant parameter.  Variations in seal behaviour were observed for bond frames of different widths, as 150\,\textmu m lines exhibited lower shear strength values compared to thinner 90\,\textmu m and 60\,\textmu m broad lines of the same height. This could be explained by wider frames having more significant Au:Sn interface surface area, thus more availability for premature gold diffusion and the formation of high-temperature intermetallic compounds. On the other hand, wider frame widths were found more robust in maintaining the hermeticity of packages. In immersion tests, all 150\,\textmu m and the majority (80\%) of 90\,\textmu m-wide frames kept enclosed copper pieces intact after 12 weeks in water. Leakages were observed commonly for thinner frames as a result of a higher probability of misalignment between lid and base dies. In thinner frames, the appearance of voids has more of a detrimental impact because of the smaller volume of the solder available for package protection. All copper pieces enclosed by 30\,\textmu m wide AuSn seal frames showed signs of exposure to water as observed by the copper colour changes (Fig.~\ref{copper}). From mechanical stability and hermeticity perspective, optimum seal geometries found out to work the best for the die sizes examined were those of middle range of widths (60\,\textmu m and 90\,\textmu m).

A structural parameter that exhibited more influence on the shear strength values was stack height. Deposited layers were higher than 10\,\textmu m to ensure the proper ratio of gold to tin is maintained. When sequentially electroplating gold and tin from two separate solutions, two distinct metallic layers of different grain sizes were achieved. Whereas gold’s layers are deposited to form smooth layers of small grains and compact microstructure, deposition of tin from methane sulfonic acid produced films of comparably more developed morphology. Micro roughness of tin, resulting from its grain size, leads to difficulties in accurate measurement of the deposited layer, thus controlling the correct eutectic ratio, especially for thin solder layers. When chosen eutectic height is tall enough, the grain size measurement error becomes insignificant, thus allowing for successful Au:Sn proportion deposition. Any premature Au:Sn intermixing and resultant consumption of metals favouring the formation of intermetallic compounds different to Au5Sn is minimised with thicker layers deposited.  All samples of as-deposited AuSn heights of 10\,\textmu m and 15\,\textmu m achieved eutectic composition, as observed by the presence of a single uniform metal phase in post-reflow SEM images and in-situ observed reflow temperature.

\begin{figure} 
    \centering
    \includegraphics[width=1\linewidth]{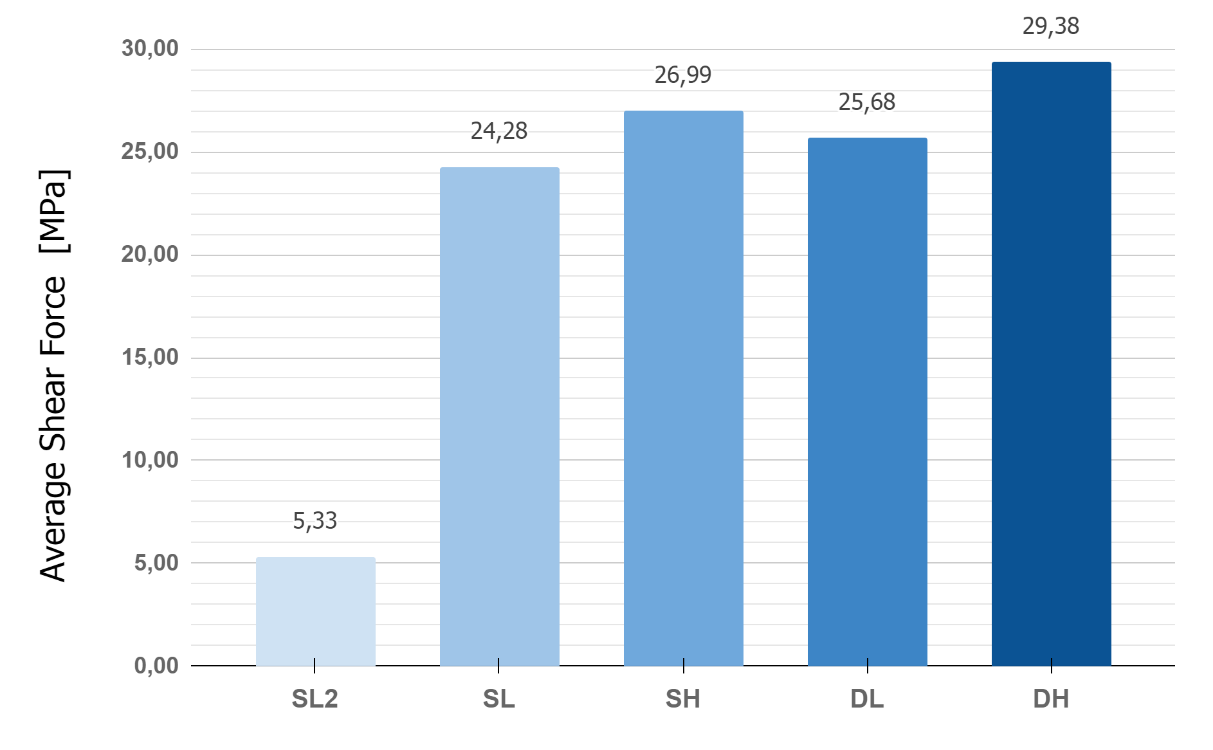}
    \caption{Average shear force for samples of different as-plated solder heights. SL-10\,\textmu m total solder on-base die only; SH-15\,\textmu m total solder on-base die only; DL-20\,\textmu m total solder on base and lid dies;  DL-\,\textmu m total solder on base and lid dies.}
    \label{asf}
\end{figure}

Four different eutectic height groups were evaluated: SL with 10\,\textmu m of as-plated eutectic height on the base die; SH with 15\,\textmu m of as-plated eutectic height on the base die; DL with 10~µm of as-plated eutectic height on both base and lid dies (total 20\,\textmu m) and  DL with 15\,\textmu m of as-plated eutectic height on both base and lip die (total of 30\,\textmu m). Additionally, the SL2 group consisting of dies with 7\,\textmu m -high AuSn eutectic on the base die with Cr adhesion layer was tested (Fig.~\ref{asf}). The latter group yielded the lowest shear strength values as the chromium layer was proven insufficient to provide an adequate level of adhesion for grown eutectic. What is more, selective etching of chromium thin film from the die surface during the last step of processing led to the premature rapid oxidation of the electroplated tin surface, further hindering subsequent bonding steps. For that reason, results from group SL2 are not included in the overall statistics of the experiment. For the remaining samples with Titanium deposited as an adhesion layer, the average shear force measured was more significant for higher thicknesses of the deposited solder. For the groups of dies with solder stack present only on one die, the variation in average shear strength between solder heights of 10 and 15\,\textmu m was only 2.71\,MPa . Doubling the amount of solder by depositing eutectic stack on both base and lid wafers led to an average shear strength increase of 1.4\,MPa for samples with double-10\,\textmu m stacks, and 2.39\,MPa for samples with double-15\,\textmu m stacks. It must be noted that when double solder height is considered, a considerably more extensive excess solder spill-out was observed for samples of all geometries. Furthermore, all double-height-bonded samples exhibited different shear fracture modes, mainly breaking at the bond interface. Considering the increased cost and complexity of double-stacked samples against only a minor increase in shear strength, it was concluded that stacks deposited on only one of the package dies were sufficient to provide a reliable seal.

\begin{figure} [h]
    \centering
    \includegraphics[width=1\linewidth]{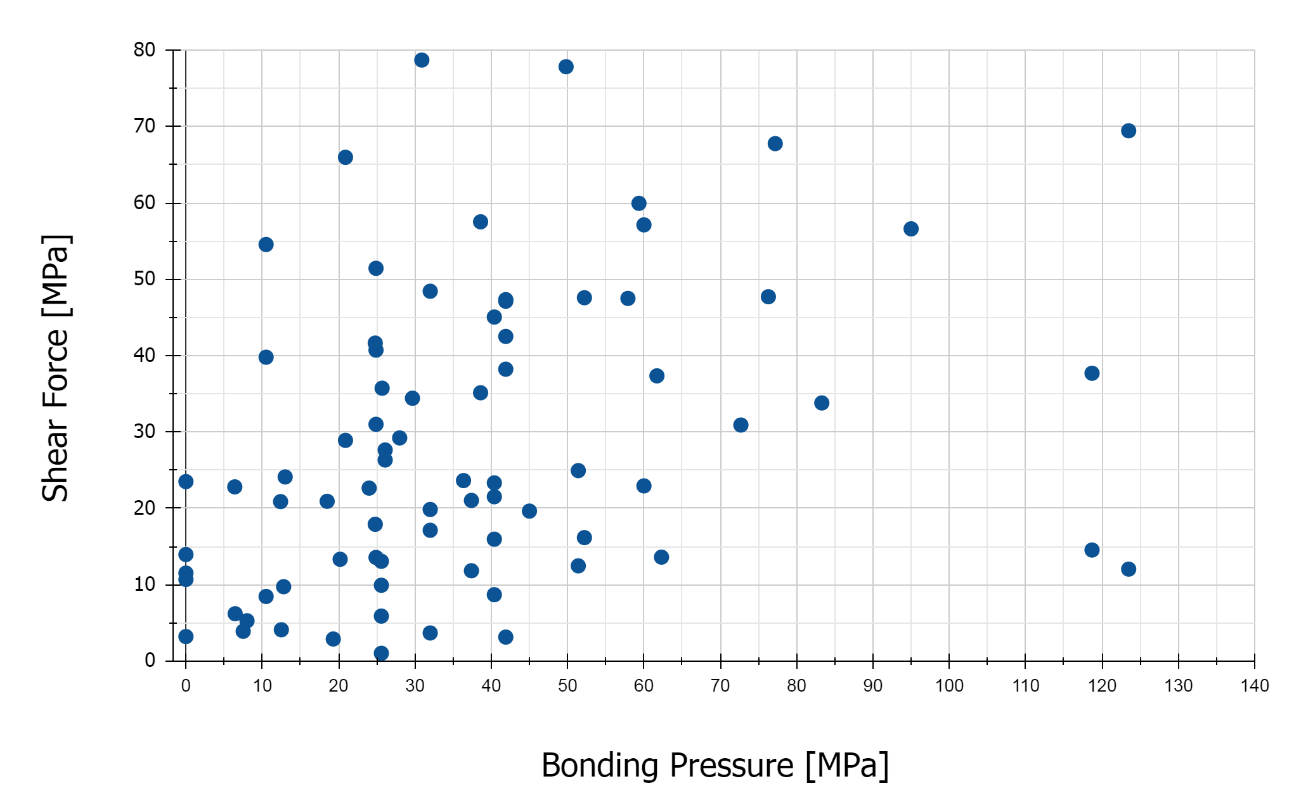}
    \caption{Measured shear force in relation to pressure applied to the sample during the bonding procedure.}
    \label{shearpress}
\end{figure}

Analysis of measured shear force against pressure applied onto the seal frames during the bonding process has shown that with an increase of bonding pressure, achievable shear strength is on average higher (Fig.~\ref{shearpress}). This is in part caused by improved mixing of solder phases during reflow under pressure. Applying pressure aids in the removal of voids as well as in the improvement of wetting of solder to the substrate; however, it causes bond line widening and increases the possibility of solder squeeze out and spillages. Very high bonding pressures were also found to cause minor shifts in dies alignment when the surface of the bonder was not perfectly levelled. On the other hand, minimal or no bonding pressure was found to be a parameter used in 40\% of all samples from the group of the weakest bonds that failed upon regular handling. Nonetheless, applying low to moderate pressures along with the suitable adjustments of other bond parameters, such as eutectic height, makes it feasible to attain shear strengths in the range above 20\,MPa, applicable for intended implantable purposes. 
\begin{figure} [h]
    \centering
    \includegraphics[width=1\linewidth]{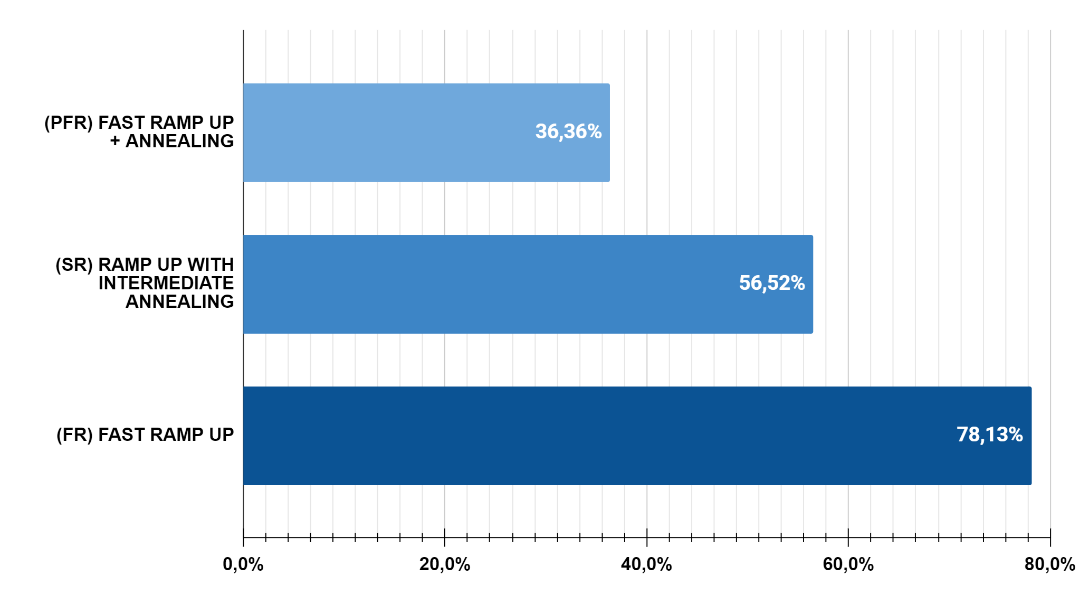}
    \caption{The average shear force between samples reflown with different temperature profiles, with or without pre-bonding annealing step.}
    \label{shearreflow}
\end{figure}
Clear differentiation between the total yield of well-bonded samples can be observed for dies reflown with different temperature profiles (Fig.~\ref{shearreflow}). Samples that were heat-treated in a vacuum oven before final reflow successfully bonded only in 36\% cases, while using the same temperature profile on samples without any prior annealing led to an increase in yield to 78\% of well-sealed samples. Typically, pre-reflow annealing is employed to increase the bondability of as-deposited layers over time by minimising the rate of diffusion of Au into Sn and the formation of unwanted intermetallics~\cite{demir2014fabrication}. However, in our case, annealing negatively impacted bonding process yield, likely due to increased Ti adhesion layer consumption. Application of intermediate annealing step at 200\degree C directly before reflow above eutectic temperature also had a negative impact on process yield. Thus preferably from the process yield perspective, reflow at a temperature above eutectic should occur as soon as possible, with as fast as possible ramp up any without any prolonged exposure to temperatures above melting temperature of Sn.

\begin{figure} 
    \centering
    \includegraphics[width=1\linewidth]{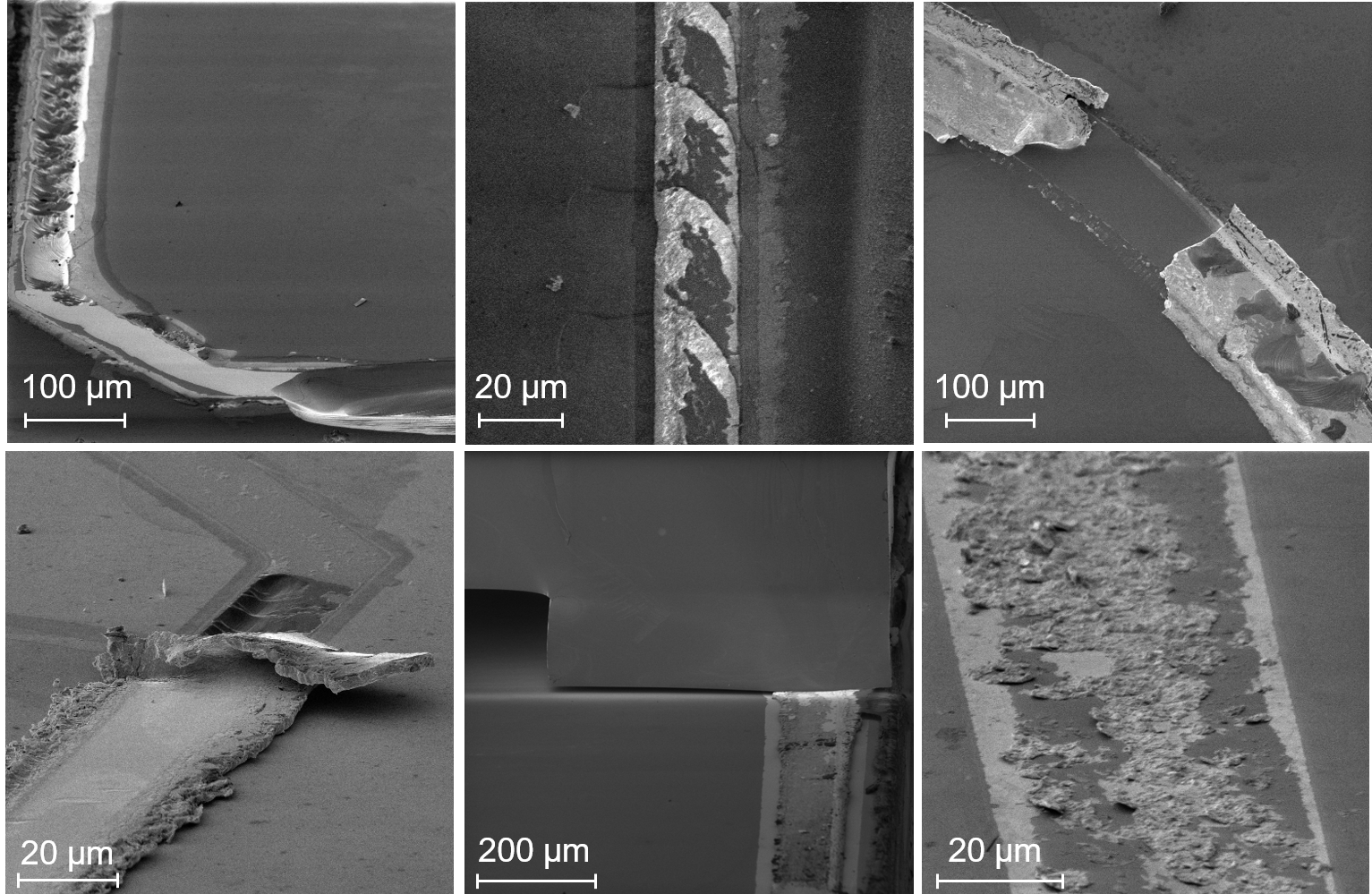}
    \caption{SEM pictures of fracture modes observed at the debonded interface of samples following shear testing.}
    \label{fracture}
\end{figure}

Fig.~\ref{fracture} represents SEM pictures of various fracture modes identified after shear testing. For a large number of samples analysed, the transfer of Ti:Au thin film bi-layer from the surface of the silicon onto the bond frame was observed. This indicates the failure occurring at the solidified solder and adhesion layer interface, pointing out the good mechanical stability and quality of the bond. The exact location for that type of failure could occur both at the interface between AuSn and Ti, as well as Ti and SiO$_2$. The high occurrence rate of this fracture mode may indicate the need for improvements of the adhesion layer to limit the dissolution of Ti into the solder. Another type of fracture identified is the one within the solid Si itself, represented as fragmentary remains of the corresponding sample from the pair attached to the solder frame. This fracture type was the second most observed and shows the bond strength surpassing fracture strengths of silicon/silicon oxide die. The least observed fracture mode was the one occurring within the bulk of the bond frame bulk itself, usually caused by either the solder imperfections such as voids or by the wrong composition of the solder and, therefore, phases non-uniformity.

\begin{figure} 
    \centering
    \includegraphics[width=1\linewidth]{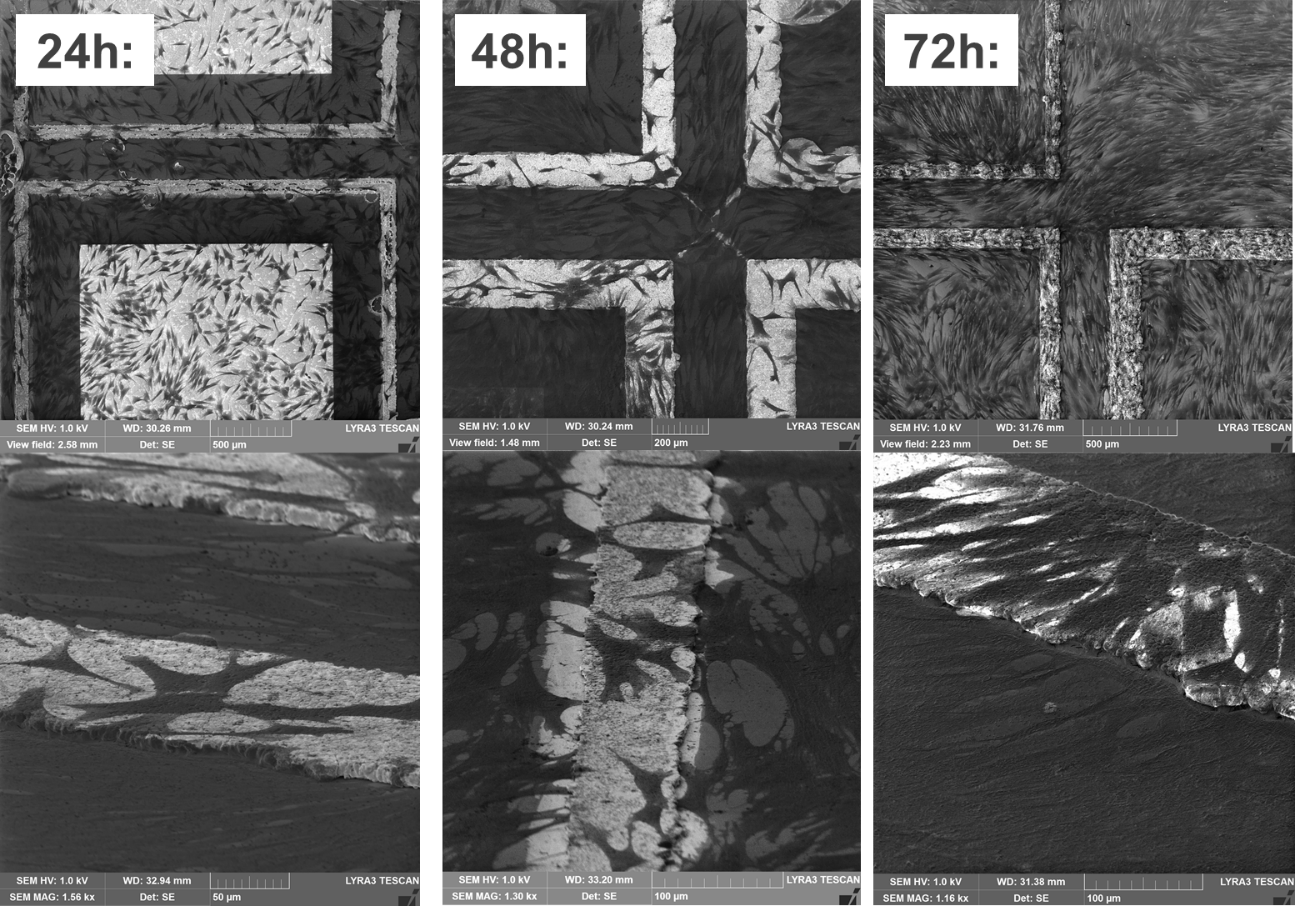}
    \caption{SEM pictures of reflown chip surface with AuSn solder lines used for fibroblast cell culturing after 24, 28, and 72 hours of cell culture.}
    \label{cells}
\end{figure}

Cell adhesion and morphology were observed by SEM on the surface of reflown AuSn eutectic after culturing for 1, 2, and 3 days. Representative surfaces after incubation in the cell culture medium are shown in (Fig.~\ref{cells}). No cytotoxicity effect on fibroblasts structure could be observed between time points of 1, 2 and 3 days, with cells consistently exhibiting normal, spindle-like morphology. Cell density exponentially increased over the course of incubation time. The continued proliferation of the fibroblast after three days has created a tissue-like structure on the surface of samples which suggest the biocompatibility and non-cytotoxicity of the substrates \cite{keshavarz2016functionalized, keshavarz2017cell}. An untreated area of the samples was considered as a control. Increasing numbers of fibroblasts attached to the surface and in contact with each other were observed. The clear preference of cellular attachment on the silicon oxide compared to the reflown AuSn was noted, with a sparser cellular layer covering solder lines. Nonetheless, fibroblasts present on the solder frames remained of healthy morphology, with only a few individual cells exhibiting round shape. After culturing for more than three days, the density of fibroblasts coverage was high enough to cover the majority of sample surfaces, proving no cytotoxic effect of sequentially electroplated AuSn solder. 

\section{Conclusions}
A method for creating solid and hermetic seals using AuSn alloy sequentially electroplated from commercially available solutions is presented.  Fabrication of eutectic frames is performed using standard microfabrication methods, and thanks to the maximum process temperature of 320\degree C can be applied to temperature-sensitive devices, such as CMOS microelectronics dies. The use of electroplating over other methods of alloy deposition allows for a parallel formation of multiple solder frames with great geometry control, while the choice of plating sequentially significantly reduces problems of plating solution stability and shelf life. The wide availability of considerably low toxicity gold and tin electroplating chemistry makes the process both cost and time effective. However, because of different Au and Sn layer morphology, it can be applied only where a thicker solder layer can be tolerated. Bonding tests with varying parameters of applied temperature, reflow profile, and frame geometry, along with immersion tests, have shown the possibility of this method to create an enclosed, mechanically robust hermetic environment.  Across all geometries and process parameters tested, the average measured bonding strength was over 28\,MPa with a bonding success yield of 73\%. The dominant shear fracture observed was at the interface of the adhesion layer, pointing to the need for improvement under solder metallisation. Nonetheless, even with only Ti/Au stack underneath, samples have exhibited excellent mechanical properties. Analysis of average bond shear force between samples having either single- or double-sided solder rings has shown no need for solder presence on both base and lid dies. This means that for applications where CMOS dies are used as a base, there would be no need for on-chip electroplating, as solder frames on separately processed lid chips would be sufficient to provide a high-quality seal. Immersion tests conducted for the duration of three months have shown no signs of moisture penetration in samples sealed with frames of 90\,\textmu m and 150\,\textmu m width, whereas the majority of thinner 30\,\textmu m-wide frames have exhibited leakage. As a result of cell culture experiments, AuSn reflown layers have not shown signs of high toxicity to the biological environment. It would be of future interest to investigate different under-layer metallisation to further improve the mechanical stability of seals.

Unquestionably one of the most significant trends for the future of implantable devices is miniaturisation. The designs of novel implantables are characterised by complicated architectures and expectations of the possibility of high yield manufacturing. Therefore, the device's design for manufacturability, including packaging methodology, should be considered from the beginning. Presented AuSn-based eutectic sealing can be applied to a variety of substrate materials and used on a larger scale, such as wafer-level, thus significantly improving the process throughput.

The presented method appears suitable packaging approach to protect integrated circuitry in implantable applications or any other application where a device must be protected from an external operating environment. The possibility to hermetically seal IC-electronics within an implant, robustly and straightforwardly, as with using AuSn eutectic, can pave the way to expand the horizon of implantable devices into the new areas. Such narrow, hermetic seals do not require any specialised microfabrication approaches, are not toxic, and provide long-term robustness.

Once combined with the right approach to the challenges of power and data transmission out of implants \cite{liu2020bidirectional}, correct packaging methodology could enable future devices to be potentially safer, less invasive, and easier to implant~\cite{ahmadi2019towards, feng2018chip}.

\ack
The authors would like to thank Matthew Cavuto for the help with the mechanical design, 3D printing, and assembly of shear testing setup;  Dr. Florent Seichepine for the general advice on processing steps; and Applied Microengineering Ltd. UK, especially Anna Draisey for process advice and valuable discussions regarding eutectic bonding. 
\newline
\newline
\textbf{References}
\newline
\bibliographystyle{unsrt}
\bibliography{bibliography.bib}

\end{document}